\providecommand{\mode}{0}

\if 0\mode
    \documentclass[twocolumn,5p,12pt,a4paper,final]{elsarticle}
\else
    \documentclass[review,preprint,12pt,a4paper]{elsarticle}
    \usepackage{lineno}
\fi
\biboptions{sort&compress}

\usepackage[colorinlistoftodos,textwidth=3cm,obeyFinal]{todonotes}

\usepackage{amssymb}
\usepackage{siunitx}
\DeclareSIUnit{\atmosphere}{atm}

\usepackage[utf8]{inputenc}
\usepackage{hyperref}
\usepackage{bm}
\usepackage{amsmath}
\usepackage[version=4]{mhchem}
\usepackage{filecontents}
\usepackage{subfig}
\usepackage{listings}

\usepackage{color}
\definecolor{gray}{rgb}{0.4,0.4,0.4}
\definecolor{darkblue}{rgb}{0.0,0.0,0.6}
\definecolor{cyan}{rgb}{0.0,0.6,0.6}

\lstdefinelanguage{XML}
{
  morestring=[b]",
  morecomment=[s]{<!--}{-->},
  commentstyle=\color{gray},
  stringstyle=\color{black},
  identifierstyle=\color{darkblue},
  keywordstyle=\color{cyan},
  morekeywords={name,type,default,mechanism,formula,A,E,n,T,state_model,s1,s2,units,multpi,label,species,enthalpy_surface,solid_conduction,pre_exp,emissivity,gsi_mechanism}
}

\newcommand{\makelistingbox}[2]{%
\newsavebox{#1}
\begin{lrbox}{#1}%
\begin{minipage}{8.5cm}%
\lstinputlisting[language=XML,frame=single]{#2}
\end{minipage}
\end{lrbox}}

\journal{SoftwareX}

\newcommand{\etal}{et al.}
\newcommand{\eref}[1]{Eq.~\ref{e:#1}}
\newcommand{\fref}[1]{Fig.~\ref{f:#1}}
\newcommand{\sref}[1]{Sec.~\ref{s:#1}}
\newcommand{\mpp}{Mutation$^{++}$}
\newcommand{\set}[1]{\ensuremath{\mathcal{#1}}}
\newcommand{\class}[1]{\texttt{#1}}

\begin{document}
%TC:ignore 
% \section*{Word counts} 
% \wordcount
% \clearpage
%TC:endignore 

\begin{frontmatter}
%TC:ignore 


\title{\mpp: MUlticomponent Thermodynamic And Transport properties for IONized gases in C++}


\author{James B. Scoggins\corref{corjb}\fnref{fnjb}}
\ead{scoggins@vki.ac.be}
\cortext[corjb]{Corresponding author.}
\fntext[fnjb]{Now at the Centre des Math\'{e}matiques Appliqu\'{e}es, \'{E}cole Polytechnique, France}

\author{Vincent Leroy\corref{}\fnref{fnvl}}
\fntext[fnvl]{Now at Rayference, Belgium}
\author{Georgios Bellas-Chatzigeorgis\corref{}}
\author{Bruno Dias\corref{}}
\author{Thierry E. Magin\corref{}}

\address{von Karman Institute for Fluid Dynamics, B-1640 Rhode-St-Gen\`{e}se, Belgium}
%TC:endignore

\begin{abstract}
The \mpp{} library provides accurate and efficient computation of physicochemical properties associated with partially ionized gases in various degrees of thermal nonequilibrium.  With v1.0.0, users can compute thermodynamic and transport properties, multiphase linearly-constrained equilibria, chemical production rates, energy transfer rates, and gas-surface interactions.  The framework is based on an object-oriented design in C++, allowing users to plug-and-play various models, algorithms, and data as necessary.  \mpp{} is available open-source under the GNU Lesser General Public License v3.0. 
\end{abstract}


\end{frontmatter}

% \if 1\mode
%     \linenumbers
% \fi


\section{Motivation and Significance}
\label{s:introduction}

The evaluation of thermochemical nonequilibrium, partially ionized gas properties is essential for a wide range of applications, including hypersonic flows, solar physics and space weather, ion thrusters, medical plasmas, combustion processes, meteor phenomena, and biomass pyrolysis.  For example, the prediction of hypersonic flow plays an important role in the development of thermal protection systems for atmospheric entry vehicles.  Such flows span a broad range of temporal scales, from local thermodynamic equilibrium to thermo-chemical nonequilibrium.  As such, myriads of physicochemical models, data, and algorithms are used in today's hypersonic Computational Fluid Dynamics (CFD) codes and represent a significant body of work in the scientific literature.  The thermochemical models employed in these codes directly affect the evaluation of gas properties necessary to close the conservation laws governing the fluid.   These include mixture thermodynamic and transport properties, species chemical production rates, and energy transfer rates.  Each of these properties further depends on the selection of a variety of specialized algorithms and data, such as species partition functions, transport collision integrals, and reaction rate coefficients.

The implementation, testing, and maintenance of the models, algorithms, and data required to simulate thermal nonequilibrium flows represent a significant cost, in terms of human resources and time necessary to develop a simulation tool.  As new models, algorithms, or data become available, additional effort is required to update existing codes, especially when models are ``hard-coded.''  A number of commercial and academic software packages are available which provide gaseous properties, including CEA~\cite{Gordon1994}, EGlib~\cite{Ern1996a}, PEGASE~\cite{Bottin1999}, Chemkin~\cite{Kee2000}, MUTATION~\cite{MaginThesis}, Cantera~\cite{Goodwin2017}, and KAPPA~\cite{Campoli2019}, however, these libraries tend to focus on a specific application, a narrow range of collisional time-scales, or are specialized in providing only certain types of properties.

These observations have led to the desire to reduce the work necessary to implement new models and algorithms and centralize their development into a single software library which may be used by multiple CFD codes to maximize code reuse, testing, and open collaboration.  This paper presents the \mpp{} library, which has been developed to meet this objective.  The library is designed with several goals in mind, including
\begin{enumerate}
\item provide accurate thermodynamic, transport, and chemical kinetic properties for multicomponent, partially ionized gases,
\item ensure the efficient evaluation of these properties using state-of-the-art, object-oriented algorithms and data structures in C++,
\item be easily extendable to incorporate new data or algorithms as they become available,
\item interface to any simulation tool based on the solution of conservation laws through a consistent and logical interface,
\item use self-documenting database formats to decrease data transcription errors and increase readability, and
\item be open source to promote code and data sharing among different research communities.
\end{enumerate}

The latest version of \mpp{} (v1.0.0) has recently been released open-source under the Lesser GNU Public License (LGPL v3) and is freely available on Github\footnote{\url{https://github.com/mutationpp/Mutationpp}}.  In the remainder of the paper, we present an overview of the library and its impact on the research community to-date.  In particular, the four main modules of the library --- thermodynamics, transport, chemical kinetics, and gas-surface interaction --- are presented, with a few examples to illustrate the library's use.

\section{Software Description}
\label{s:software}

\subsection{Generalized Conservation Equations}
While it is beyond the scope of this article to describe in detail all the various physicochemical models that are present in the literature, it is useful to briefly present a generalized model which has been used in the design of the library.  For a more complete discussion, see the work of Scoggins \cite{ScogginsThesis-Chap2}.  We consider a generalized conservation law of the form,
\begin{equation}
\partial_t \bm{U} + \nabla_{\bm{x}} \bm{F} = \bm{S},
\label{e:general-conservation}
\end{equation}
where
$\bm{U} = \begin{bmatrix}
    \tilde{\rho}_i & \rho \bm{u} & \rho E & \rho \tilde{e}^m
\end{bmatrix}^T$,
is a vector of species mass, momentum, and total and internal energy densities, $\bm{F}(\bm{U}, \nabla_{\bm{x}}\bm{U})$ represents their flux, and $\bm{S}(\bm{U})$ is a source function.  The tilde over the indexed variables in the density vector denotes that these quantities must be expanded over their indices.  
The exact forms of $\bm{U}$, $\bm{F}$, and $\bm{S}$ depend on 1) the coordinate system, 2) the physical model (i.e.: Euler, Navier-Stokes), and 3) the thermochemical model of the gas (i.e.: equilibrium, reacting, multi-temperature, state-to-state).  

\begin{figure}[t!]
    \centering
    \includegraphics[width=0.48\textwidth]{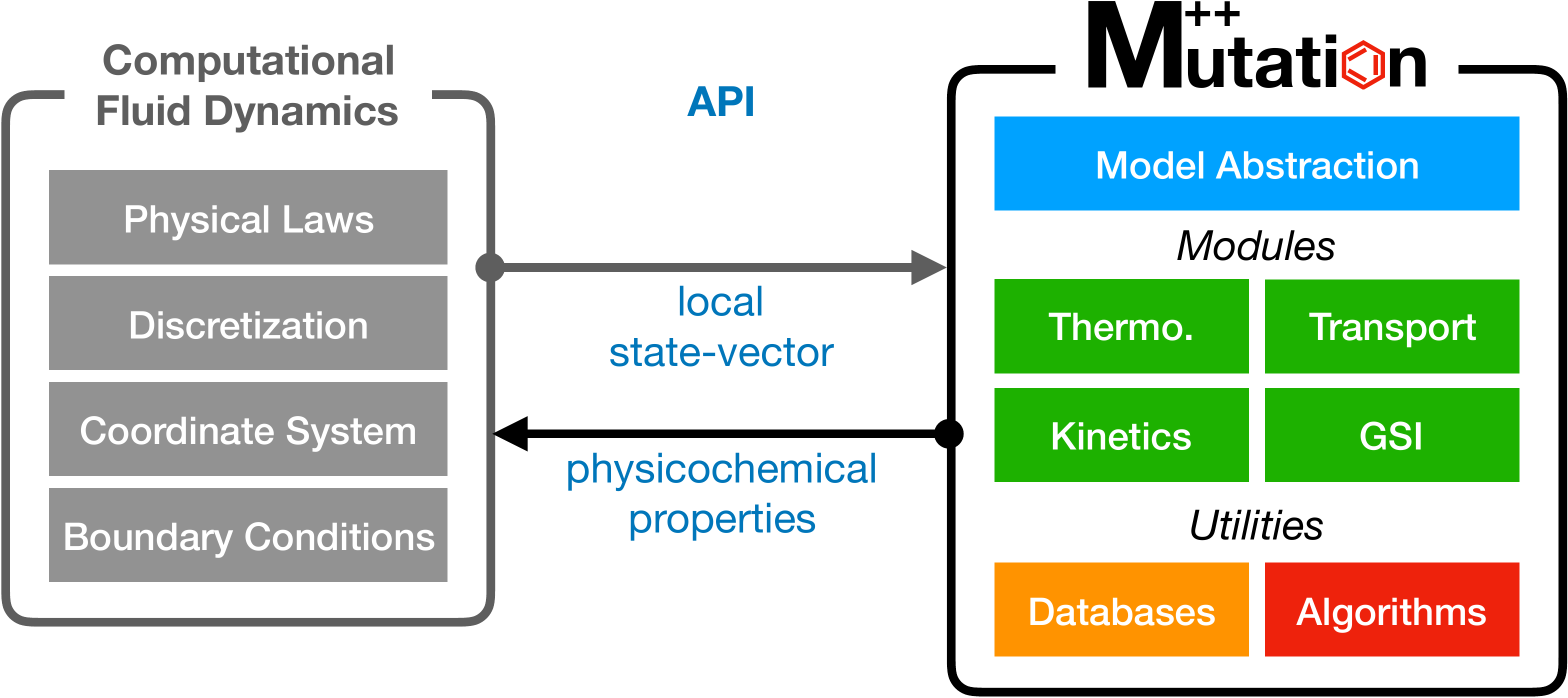}
    \caption{Overview of the \mpp{} library and its coupling to CFD.}
    \label{f:overview}
\end{figure}

We define the thermochemical state-vector as
$\hat{\bm{U}} = \begin{bmatrix}
    \tilde{\rho}_i & \rho e & \rho \tilde{e}^m
\end{bmatrix}^T$
where $\rho e = \rho E - \rho \bm{u}\cdot\bm{u} / 2$ is the static energy density of the gas.  The flux and source functions are closed by constitutive relations for thermodynamic, transport, and chemical properties of the gas.  These include quantities such as pressure, enthalpy, viscosity, thermal conductivity, diffusion coefficients, chemical production rates, and energy transfer source terms.  In general, these properties are only functions of the local state-vector $\hat{\bm{U}}$ and possibly its gradient.  This fact allows us to separate the solution of \eref{general-conservation} into two separate domains with limited coupling controlled by the CFD solver and \mpp, as shown in \fref{overview}.

\subsection{Software Architecture}

\mpp{} is designed with a strong focus on Object-Oriented Programming (OOP) patterns in C++.  The library's Application Programming Interface (API) is thoroughly documented using the Doxygen format.  A continuous integration strategy has been employed.  Regression and black box testing are performed through a combination of the Catch2 header-only testing framework and CTest.
The primary access to the library is through a \class{Mixture} object, which is implemented as a set of submodules encapsulating clearly separated physical quantities as depicted in the simplified Unified Modeling Language (UML) diagram in \fref{uml}.
\begin{figure*}[t]
    \centering
    \includegraphics[width=0.95\textwidth]{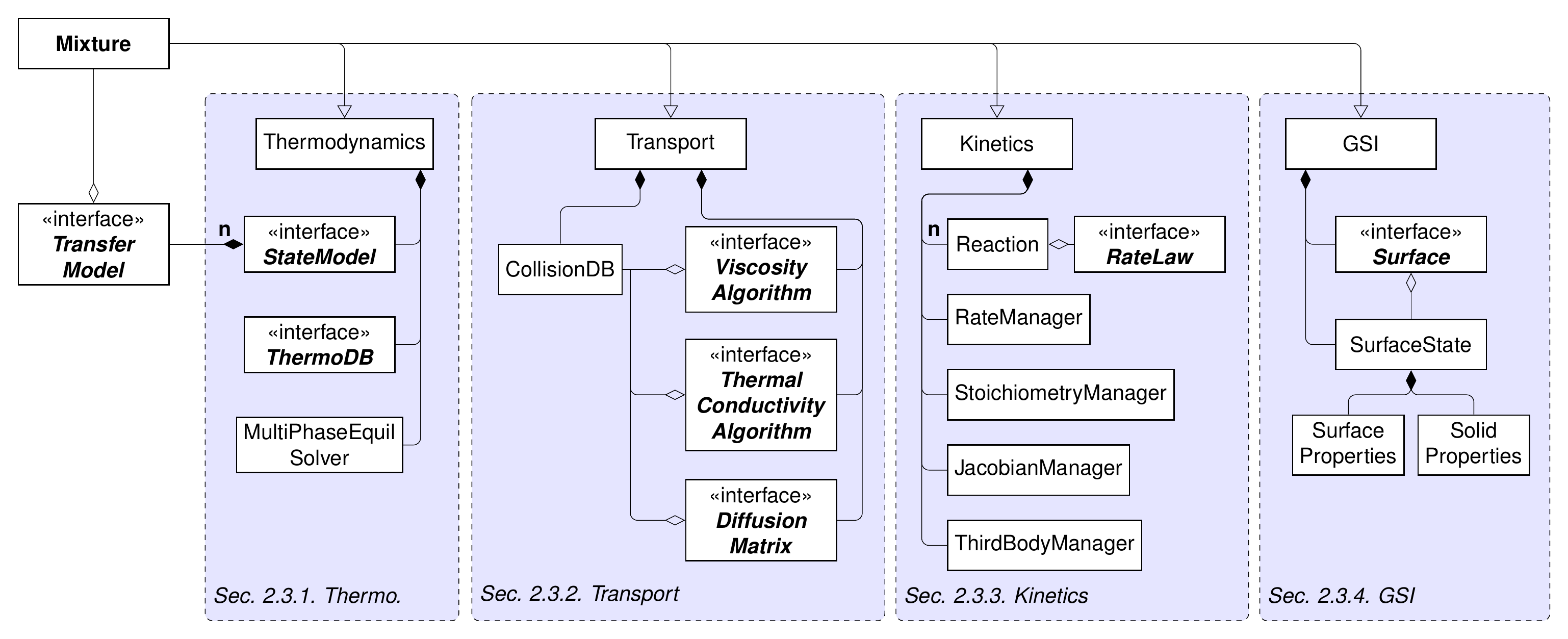}
    \caption{Simplified UML class diagram showing major components of the \mpp{} architecture.}
    \label{f:uml}
\end{figure*}

\subsection{Software Functionalities}

Each module in \fref{uml} is described in the following subsections.  Specific examples of some of the outputs that the library can provide are given in the \sref{examples}.

% \begin{figure*}[t]
%     \centering
%     \includegraphics[width=0.8\textwidth]{figures/input_files.pdf}
%     \caption{Listing of input files associated with a CHO mixture.}
%     \label{f:input-files}
% \end{figure*}

\makelistingbox{\mixturefile}{mixture.xml}
\makelistingbox{\mechanismfile}{mechanism.xml}
\makelistingbox{\collisionfile}{collisions.xml}
\makelistingbox{\gsifile}{gsi.xml}

\begin{figure*}[t]
    \centering
    \begin{minipage}{0.45\textwidth}
        \subfloat[Mixture file.]{\usebox{\mixturefile}\label{f:mixture-file}}\\
        \subfloat[Collision database.]{\usebox{\collisionfile}\label{f:collision-file}}
    \end{minipage}\hspace*{1cm}
    \begin{minipage}{0.45\textwidth}
        \subfloat[Reaction mechanism.]{\usebox{\mechanismfile}\label{f:mechanism-file}}\\
        \subfloat[GSI mechanism.]{\usebox{\gsifile}\label{f:gsi-file}}
    \end{minipage}    
    \caption{Example input files defining a mixture model.}
    \label{f:input-files}
\end{figure*}

\subsubsection{Thermodynamics}
\label{s:thermo}

The thermodynamics module provides pure species and mixture thermodynamic quantities, such as enthalpy, entropy, specific heats, or Gibbs free energies.  Mixture thermodynamic quantities are derived as sums of pure species properties, weighted by the composition of the mixture.  Thermodynamic data for pure species can be found in several references \cite{Ruscic2004,Ruscic2005,Burcat2005,Ruscic2013,Blanquart2007,Blanquart2009,Narayanaswamy2010,Blanquart2015,Goldsmith2012}.  Differences exist between each database, such as their format, temperature range of applicability, or degree of nonequilibrium supported.  Such differences often sway simulation tool designers to select a single database format to support, or hard-code thermodynamic data directly into their models.  This approach makes it difficult to update data as needed, or compare with other tools using different databases.

The \mpp{} framework provides an abstraction layer which enforces a weak coupling between the concrete thermodynamic database, used for any given set of species, and the computation of mixture thermodynamic quantities.  Such a design provides the flexibility to swap out different databases as needed, with minimal effort.  The NASA 7- and 9-coefficient polynomial databases \cite{McBride1993,McBride1993a,Gordon1999,McBride2002} and a custom XML format which implements a Rigid-Rotor/Harmonic-Oscillator (RRHO) model are currently implemented.  The NASA format is widely used and provides thermodynamic properties for pure species in thermal equilibrium.  The RRHO model is suitable for thermal nonequilibrium calculations. 
In addition, a new database including more than 1200 neutral and ionized species containing \ce{C}, \ce{H}, \ce{O}, and \ce{N} is shipped with the library in the NASA 9-coefficient format.  The details of this database have been published in \cite{Scoggins2017}.  The user can specify the concrete thermodynamic model when creating a mixture (\fref{mixture-file}).

A closely related task to the calculation of thermodynamic properties is the solution of chemical equilibrium compositions.  The efficient and robust computation of multiphase, constrained equilibrium compositions is an important topic in several fields, including combustion, aerospace and (bio)chemical engineering, metallurgy, paper processes, and the design of thermal protection systems for atmospheric entry vehicles (e.g., \cite{Chan1992, Pajarre2008, Koukkari2011, Milos1997, Howard2012, Rabinovitch2014}).  Several challenges associated with computing chemical equilibria make conventional methods hard to converge under some conditions \cite{Reynolds1986,Gordon1994,McBride1996,Bishnu1997, Bishnu2001}.  A new multiphase equilibrium solver, based on the single-phase Gibbs function continuation method \cite{Pope2003,Pope2004}, has been developed specifically for \mpp.  The Multiphase Gibbs Function Continuation (MPGFC) solver is robust for all well-posed constraints.  More details about the solver can be found in \cite{Scoggins2015b}.

\subsubsection{Transport}

Closure of transport fluxes is achieved through a multiscale Chapman-Enskog perturbative solution of the Boltzmann equation, yielding asymptotic expressions for the necessary transport coefficients, such as thermal conductivity, viscosity, and diffusion coefficients \cite{Ferziger1972,Mitchner1973,Giovangigli1999,Graille2009,Nagnibeda2009}.  Explicit expressions for these coefficients are derived in terms of linear transport systems through a Laguerre-Sonine polynomial approximation of the Enskog expansion at increasing orders of accuracy \cite{Magin2004, Magin2004a, Scoggins2016}.  These linear systems are functions of transport collision integrals and the local state-vector, and may be solved through a variety of methods.

Collision integrals represent Maxwellian averages of collision cross-sections for each pair of species considered in a given mixture \cite{Hirschfelder1954,Ferziger1972}, weighted depending on the Laguerre-Sonine polynomial order used \cite{Chapman1998}.  The preferred method to compute collision integrals is to numerically integrate from accurate \textit{ab initio}~potential energy surfaces.  Such data is available for several important collision systems \cite{Wright2005,Wright2007, Bruno2010}.  When potential energy surfaces are not available, collision integrals are integrated from model interaction potentials.  The evaluation of these model integrals can be partitioned in several ways ---  neutral-neutral, ion-neutral, electron-neutral, and charged interactions, heavy, electron-heavy, and electron interactions --- each with different functional dependencies on temperature and degree of ionization.  \mpp{} introduces a custom XML format for storing collision integral data (\fref{collision-file}) which
\begin{enumerate}
\item is self-documenting,
\item is easily extensible in data and model type,
\item provides customizable default behavior for missing data, and
\item enforces consistency of standard ratios.
\end{enumerate}
Just-in-time loading and efficient evaluation of this data is handled by a \class{CollisionDB} object, as shown in \fref{uml}.

The solution of the linear transport systems represents a significant CPU time for some CFD applications.  Several algorithms have been proposed in the literature for reducing this cost \cite{Giovangigli1991,Ern1995,Ern1997,Magin2004a,Giovangigli2010}.
\mpp{} provides plug-and-play transport algorithms through the use of self-registering algorithm classes.  For example, the abstract class \class{ThermalConductivityAlgorithm}, shown in \fref{uml}, provides the necessary interface that all thermal conductivity algorithms must include, namely functions for computing the thermal conductivity and thermal diffusion ratios.  Specific algorithms are then implemented by creating a concrete class which implements the interface.  This pattern has been used for the calculation of the multicomponent diffusion matrix and shear viscosity as well. 

\subsubsection{Kinetics}

The goal of the chemical kinetics module is the efficient and robust computation of species production rates due to finite-rate chemical reactions.  For reaction set $\set{R}$ involving species in $\set{S}$, we consider production rates of the form
\begin{equation}
\frac{\dot{\omega}_k}{M_k}= \sum_{r\in\set{R}} \Delta \nu_{kr} \bigg[k_{fr}\prod_{j\in\set{S}}\tilde{\rho}_j^{\nu^{'}_{jr}}-k_{br}\prod_{j\in\set{S}}\tilde{\rho}_j^{\nu^{''}_{jr}}\bigg] \Theta_r,
\label{e:species-prodrate}
\end{equation}
where a full description of each term is given in \cite{ScogginsThesis-Chap2}. The forward reaction rate is assumed to be a function of a single, reaction-dependent temperature $k_{fr} = k_{fr}(T_{fr})$ and the backward rate is determined from equilibrium as $\smash{k_{br}(T_{br}) = k_{fr}(T_{br}) / K_{\text{eq},r}(T_{br})}$ where $T_{br}$ is a reaction-dependent temperature for the backward rate.

Apart from the reaction rate temperatures, knowledge of reaction types is essential in some energy exchange mechanisms.  Manually inputting the type of every reaction in a mechanism of hundreds or thousands of reactions can be a tedious and error-prone process.  Therefore, \mpp{} provides a unique feature which determines the type of reaction automatically when a mechanism is loaded.  The problem is formulated as classification tree \cite{Loh2011}, which can be constructed automatically using simple characteristics of each reaction.  An example of such a classification tree is provided in \fref{reaction-classification}.

\begin{figure*}[t!]
\centering
\begin{tabular}{m{9.5cm}m{3.5cm}}
    \includegraphics[scale=0.52]{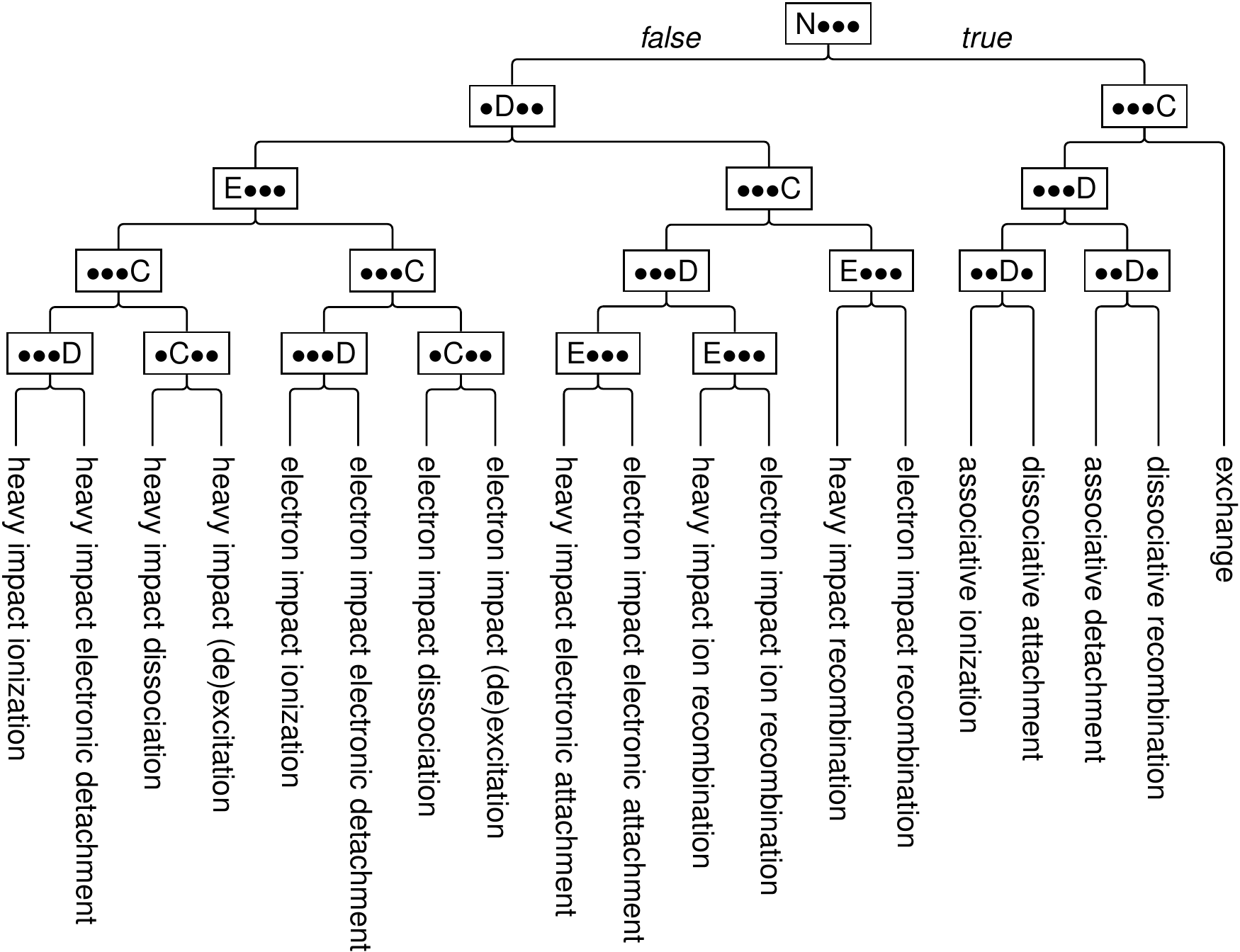}
    &
    \includegraphics[scale=0.52]{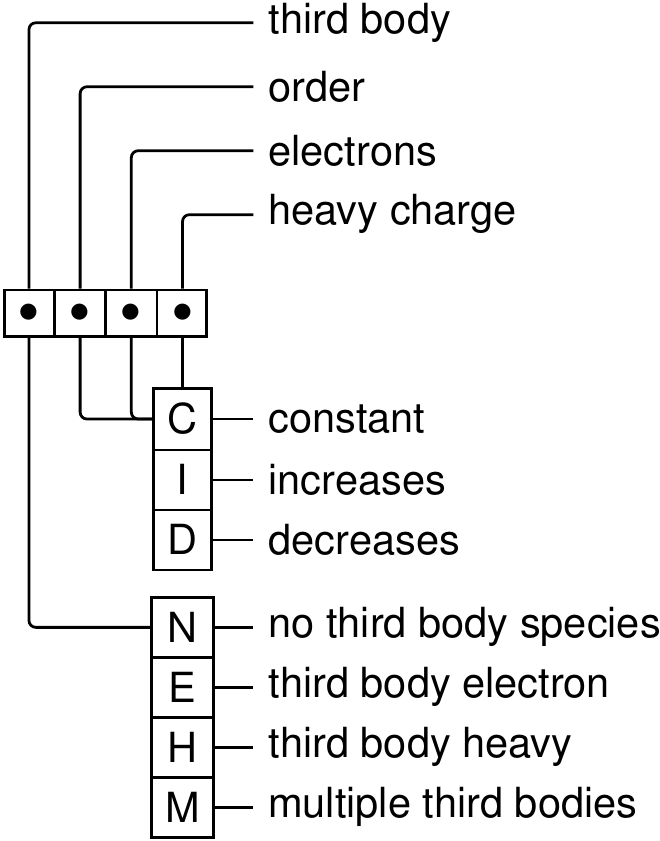}\\
\end{tabular}
\caption{Decision tree used in \mpp{} for automatically classifying chemical reactions.}
\label{f:reaction-classification}
\end{figure*}

In principle, the evaluation of \eref{species-prodrate} is straight-forward, though great care is required to do it robustly and efficiently.  A simplified class diagram of the kinetics module is presented in \fref{uml}.  The module contains a list of \class{Reaction} objects provided by the user through an XML reaction mechanism file (see \fref{mechanism-file}).  The rest of the module is comprised of a set of computational ``managers'', which are responsible for the efficient evaluation of individual parts of \eref{species-prodrate}.  These include the evaluation of reaction rates, operations associated with the reaction stoichiometry (the sum and products in \eref{species-prodrate}), and the evaluation of the third-body term, $\Theta_r$.  An additional manager class is responsible for evaluating the Jacobian of species production rates, necessary for implicit time-stepping CFD algorithms.  Finally, the \class{Kinetics} class orchestrates the use of each of these managers to evaluate \eref{species-prodrate} and its Jacobian with respect to species densities and temperatures.

\subsubsection{Gas-Surface Interactions}

The Gas-Surface Interaction (GSI) module provides surface boundary conditions for \eref{general-conservation}. They are obtained by applying the conservation of mass, momentum, and energy in a thin control-volume on a surface at steady-state in the form
\begin{equation}
\left(\bm{F}_{\text{g}} - \bm{F}_{\text{b}} \right) \cdot \bm{n} = \bm{S}_{\text{s}},
\label{e:gsi-balances}
\end{equation}
where $\bm{F}_{\text{g}}$ and $\bm{F}_{\text{b}}$ are gas and bulk phase fluxes, $\bm{n}$ is the surface normal, and $\bm{S}_{\text{s}}$ is the source term associated with surface processes.

\mpp{} provides several built-in terms that can be mixed to create a model through a custom XML format, as shown in \fref{gsi-file}. Once a model is specified, balance equations are created dynamically through an object-oriented approach forming a \class{Surface} object (\fref{uml}). The resulting nonlinear equations are functions of the surface thermodynamic state (stored as $\hat{\bm{U}} = \begin{bmatrix} \tilde{\rho}_i & \rho e & \rho \tilde{e}^m \end{bmatrix}^T$ in \class{SurfaceState}), thermochemical properties of the interface, and the two connecting phases (e.g. \class{SurfaceProperties} and \class{SolidProperties}). This framework provides flexibility for the description of a variety of surfaces (e.g. chemically active, impermeable, porous with fixed outgassing, etc.).

For each type of surface, \mpp{} provides the fluxes and source terms expressed in \eref{gsi-balances} to the client code. A very unique feature of the library is that it, on demand, solves the steady-state balances in a robust and efficient way, to obtain the boundary condition necessary for the CFD or material solvers. More information about the GSI models available can be found in \cite{BellasChatzigeorgis2018}.

\section{Illustrative examples}
\label{s:examples}

\subsection{Thermodynamic and Transport Properties}

\fref{air-properties} presents a selection of thermodynamic and transport properties computed by \mpp{} for an 11-species, isobaric air mixture in thermochemical equilibrium.  The equilibrium mole fractions and thermodynamic properties are provided using both the NASA-9 and RRHO thermodynamic databases.  Where possible, comparisons with equilibrium air curve-fits of D'Angola \etal~\cite{DAngola2008} and thermal conductivity data from Murphy \cite{Murphy1995} and Azinovsky \etal~\cite{Asinovsky1971} are also shown.  Both ``frozen'' and ``equilibrium'' curves are shown for the specific heat at constant pressure and specific heat ratio. The frozen curves neglect the dependence of the species composition on the temperature through the equilibrium reactions, while equilibrium curves do not.  This distinction is important, as it shows that these properties can vary substantially, depending on the thermochemical model employed by the user.  For more information regarding the models, data, algorithms, and interpretation of these figures, please see the discussion in \cite{ScogginsThesis-Chap4}.

\begin{figure*}[ht!]
\centering
\includegraphics[width=0.95\textwidth]{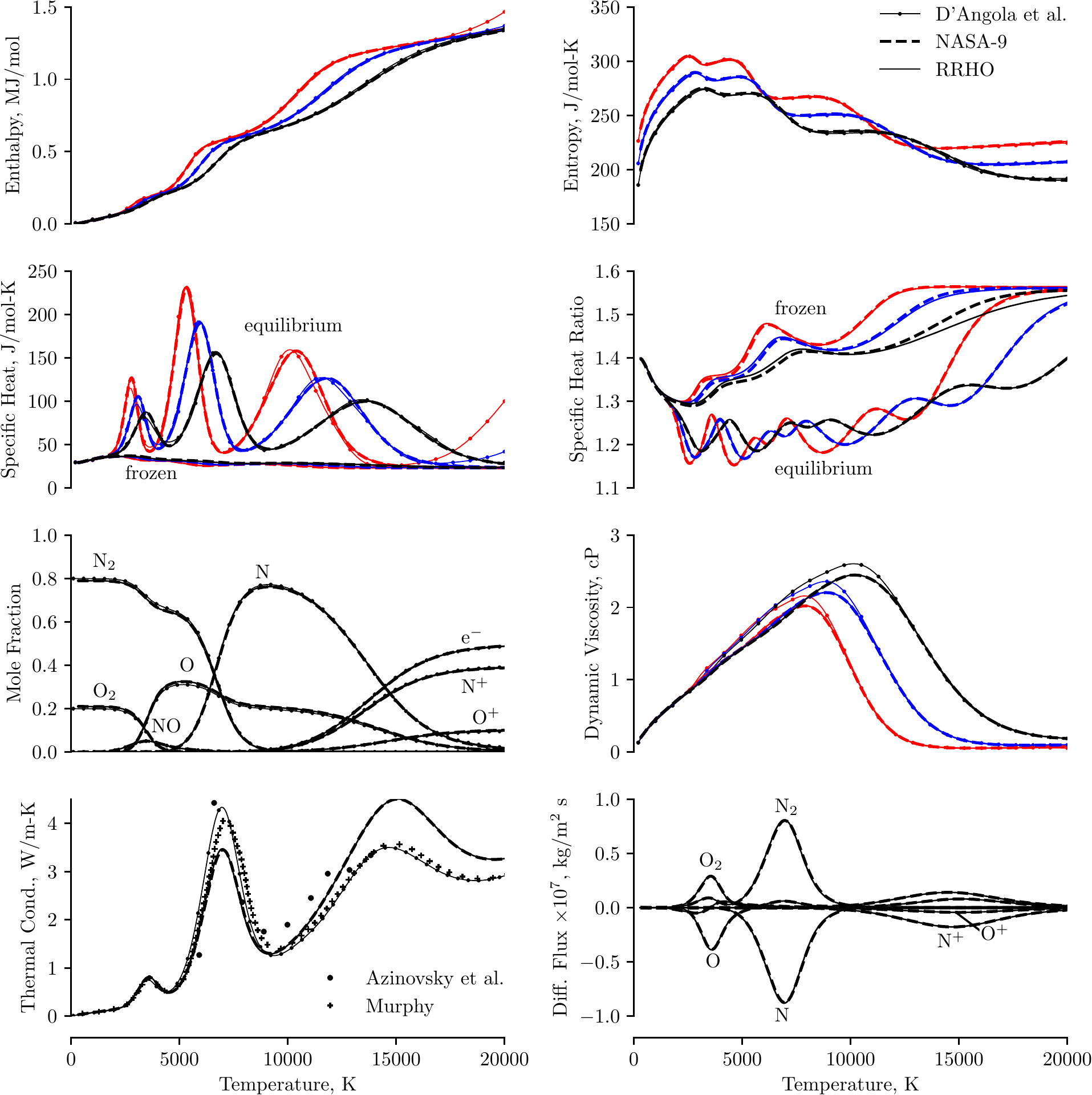}
\caption{A selection of thermodynamic and transport properties computed with \mpp{} for an isobaric air mixture in equilibrium at \SI{0.01}{\atmosphere} (red), \SI{0.1}{\atmosphere} (blue), and \SI{1}{\atmosphere} (black).}
\label{f:air-properties}
\end{figure*}

\subsection{Equilibrium Ablation Rates}

An important problem in the prediction of material response for thermal protection systems (TPS) of atmospheric entry vehicles is the solution of so-called ``B-prime'' tables.  B-prime tables describe the equilibrium gas composition at the surface of an ablating TPS as well as its mass loss rate due to reactions at the surface, such as oxidation or nitridation.  Assuming a thin control volume over an ablating TPS with equal diffusion coefficients for each species, conservation of elements inside the control volume yields
\begin{equation}
y_{w} = \frac{B'_c\;y_{c} + B'_g\;y_{g} + y_{e}}{B'_c + B'_g + 1},
\label{e:bprime}
\end{equation}
where $y$ is the elemental mass fraction for any element in the mixture, the subscripts $w$, $c$, $g$, and $e$ refer to wall, char, pyrolysis gas, and boundary layer edge properties respectively, $B' \equiv \dot{m}/(\rho_e u_e C_M)$ is a mass blowing rate, nondimensionalized by the boundary layer edge mass flux, and $C_M$ is the local Stanton number for mass transfer.  When coupled with the minimization of Gibbs energy at a known surface condition, \eref{bprime} can be solved to obtain species composition and char mass blowing rate $B'_c$.  \fref{bprime} shows such a calculation performed using \mpp{} with the custom thermodynamic database discussed in \sref{thermo}.  The $B'_c$ results are compared with those obtained with the thermodynamic database from the NASA Chemical Equilibrium with Applications (CEA) code.  More discussion on the differences in these databases can be found in \cite{Scoggins2017}.

\begin{figure}[ht!]
\centering
\includegraphics[width=0.48\textwidth]{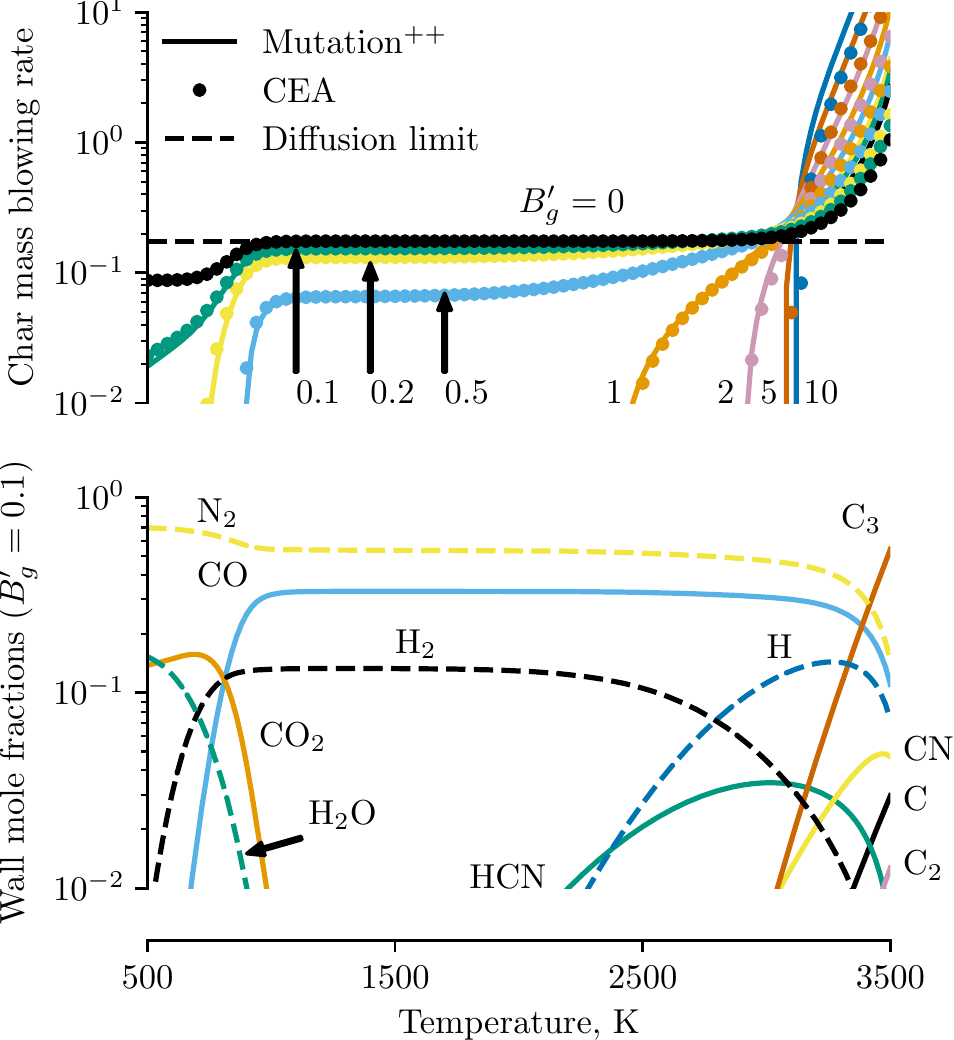}
\caption{Equilibrium char mass blowing rate and wall mole fractions for a carbon-phenolic ablator in air computed with \mpp{} \cite{Scoggins2017}.}
\label{f:bprime}
\end{figure}

\section{Impact}
\label{s:impact}

The main goal of \mpp{} is to promote open collaboration between different research groups and communities working in the broad areas of hypersonics, combustion, and plasma physics, by lowering the cost of developing new physicochemical models, data, and algorithms, for modeling gas and gas-surface phenomena.  With an efficient and extensible framework, \mpp{} can be easily coupled with existing CFD tools, allowing researchers to test effectively new thermodynamic, transport, or chemical models, physicochemical data, or numerical algorithms.  In addition, users of the library can benefit from the work of others through collaborative testing, bug fixing, and maintenance which is supported by a continuous development and integration strategy.

Since its creation, \mpp{} has been used in many diverse applications, branching out from the original motivation of hypersonic flows for atmospheric entry \cite{delValBenitez2015,Bellas-Chatzigeorgis2015,Bellemans2015, Fossati2019}.  These include the study of biomass pyrolysis \cite{Lachaud2017}, solar physics, magnetized transport \cite{Scoggins2016}, and meteor phenomena \cite{Dias2015,Dias2016}.  Application of the library to fields other than originally intended serves to highlight the extensibility of the framework and the impact it can have on basic research.  The library has also been used in limited commercial settings.  Recently, \mpp{} has been coupled to the material response codes Amaryllis of LMS Samcef (Siemens) and the Porous material Analysis Toolbox (PATO), developed jointly between C la Vie and NASA Ames Research Center.

\section{Conclusions}

The \mpp{} library provides an OOP framework for computing thermodynamic, transport, and kinetic properties of non to fully ionized gas mixtures at all points on thermochemical nonequilibrium spectrum.  The library leverages the simple dependence of thermochemical properties on the local thermodynamic state of the gas to implement a weak coupling between the computation of those properties and the simulation tools which need them, through a clean and consistent API.  

The code is freely available on GitHub with an LGPL v3.0 license, following a continuous integration development strategy with periodic versioning to alleviate backward compatibility concerns for its users.  This paper marks version v1.0.0 for the library.  Future versions will aim to provide additional features and greater flexibility for the end-user.

\section*{Acknowledgements}

The development of \mpp{} v1.0.0 was partially supported under the European Research Council Starting Grant \#259354 and Proof of Concept Grant \#713726.  The authors would like to thank Drs. Nagi N. Mansour, Jean Lachaud, and Alessandro Turchi for invaluable discussions and guidance during development of the library.  Other contributions, including development and testing of the library, are listed on the project's GitHub site.



%% References:
%TC:ignore 
\bibliographystyle{elsarticle-num}
\bibliography{biblio}

\section*{Required Metadata}

\section*{Current code version}

\begin{table}[!h]
\begin{tabular}{|l|p{6.5cm}|p{6.5cm}|}
\hline
\textbf{Nr.} & \textbf{Code metadata description} & \textbf{Please fill in this column} \\
\hline
C1 & Current code version & v1.0.0 \\
\hline
C2 & Permanent link to code/repository used for this code version & https://github.com/mutationpp/Mutationpp \\
\hline
C3 & Legal Code License   & LGPL-3.0 \\
\hline
C4 & Code versioning system used & git \\
\hline
C5 & Software code languages, tools, and services used & C++, Python, Fortran, CMake, Eigen, Catch2, Doxygen \\
\hline
C6 & Compilation requirements, operating environments \& dependencies & Linux, Mac OS X \\
\hline
C7 & If available Link to developer documentation/manual & See Github project for documentation.  \\
\hline
C8 & Support email for questions & scoggins@vki.ac.be \\
\hline
\end{tabular}
\caption{Code metadata (mandatory)}
\end{table}

%TC:endignore 
\end{document}